\documentclass[a4paper,runningheads]{llncs}

\usepackage{imakeidx}
\makeindex
\indexsetup{noclearpage}

\usepackage{hyperref}
\usepackage{times}
\usepackage{ifthen}
\usepackage{calc}
\usepackage{graphicx}
\usepackage[dvipsnames]{xcolor}
\usepackage{amssymb}
\usepackage{amsmath}
\usepackage{stmaryrd}
\usepackage{amsthm}
\usepackage{yfonts}

\usepackage{laws}
\usepackage{SETSS}
\renewcommand{\id}{\mathop{\mathsf{id}}}
\newcommand{\draftonly}[1]{}

\newif\ifarXiv
\arXivfalse

\newcommand{\localise}{localise}
\newcommand{\localised}{localised}
\newcommand{\localising}{localising}
\newcommand{\localisation}{localisation}
\newcommand{\Localisation}{Localisation}

\renewcommand{\hide}{E}
\newcommand{\Hide}[2]{{\mathop{\hide_{#1}}#2}}
\newcommand{\HideFirst}[2]{{\mathop{\hide^{\mathsf{first}}_{#1}}#2}}
\newcommand{\HideLast}[2]{{\mathop{\hide^{\mathsf{last}}_{#1}}#2}}

\newcommand{\hideuniv}{A}
\newcommand{\HideUniv}[2]{{\mathop{\hideuniv_{#1}} #2}}

\newcommand{\Complement}[1]{\overline{#1}}

\newcommand{\Keyword}[1]{\mathbf{#1}}
\newcommand{\Local}[2]{(\mathop{\Keyword{local}} #1 \spot #2)}
\newcommand{\variable}[2]{(\Var #1 \spot #2)}
\newcommand{\eguard}[1]{\Demand #1}

\newcommand{\pset}{\mathbb{P}}

\newcommand{\AtomicSteps}{\mathcal{A}}
\newcommand{\Commands}{\mathcal{C}}
\newcommand{\PCommands}{{\cal P}}
\newcommand{\ECommands}{{\cal E}}
\newcommand{\Tests}{\mathcal{T}}

\newcommand{\testsym}{\tau}
\newcommand{\programstep}{\pi}
\newcommand{\envstep}{\epsilon}

\renewcommand{\Magic}{\bot}
\renewcommand{\Abort}{\top}
\newcommand{\meet}{\sqcap}
\newcommand{\Meet}{\bigsqcap}
\newcommand{\join}{\sqcup}
\renewcommand{\Join}{\bigsqcup}

\renewcommand{\nondet}{\join}
\renewcommand{\Nondet}{\Join}
\renewcommand{\refsto}{\geq}
\newcommand{\OLDrefsto}{\mathrel{\sqsubseteq}}

\usepackage[disable,textwidth=2cm]{todonotes}
\newcommand{\backgroundintensity}{50}
\newcommand\notesb[4]{
  \todo[linecolor=red,backgroundcolor=#2!\backgroundintensity,size=\small]
  {#1: #3}{#4}
}
\newcommand{\ihsb}[2]{\notesb{IH}{green}{#1}{#2}}

\newcommand\notein[3]
{\todo[inline,linecolor=red,backgroundcolor=#2!\backgroundintensity]
  {#1 says: #3}
}
\newcommand{\ihin}[1]{\notein{IH}{green}{#1}}

\newcommand{\lmin}[1]{\notein{LM}{yellow}{#1}}

\newcommand\todobig[4]{\todo[inline,backgroundcolor=#2!\backgroundintensity,caption={#1 says: #3}]{ 
\begin{minipage}{\textwidth-4pt}#1 says: #3.\par #4\end{minipage}}}

\newcommand{\lmbig}[2]{\todobig{LM}{yellow}{#1}{#2}}

\newcounter{hours}
\newcounter{minutes}
\newcommand{\printtime}{%
  \ifthenelse{\value{hours}<10}{0}{}\thehours:%
  \ifthenelse{\value{minutes}<10}{0}{}\theminutes}

\begin{document}

\newbox{\MyDate}
\savebox{\MyDate}{\draftonly{ (\today\ \printtime)}}
\title{Handling localisation in rely/guarantee concurrency:\\ An algebraic approach\ifarXiv:\\ extended version with proofs\fi%
\thanks{This research was supported by
Discovery Grant DP19010214 from the Australian Research Council (ARC).}
}
\author{Larissa A. Meinicke \and Ian J. Hayes}
\titlerunning{Localisation in rely/guarantee concurrency\usebox{\MyDate}}
\authorrunning{L. A. Meinicke and I. J. Hayes\usebox{\MyDate}}
\institute{
The University of Queensland, Brisbane, Queensland 4072, Australia
  \draftonly{\\\vspace*{2ex} \today~\printtime}
}

\maketitle

\begin{abstract}
The rely/guarantee approach of Jones extends Hoare logic with rely and guarantee conditions
in order to allow compositional reasoning about shared-variable concurrent programs. 
This paper focuses on localisation in the context of rely/guarantee concurrency 
in order to support local variables.
Because we allow the body of a local variable block to contain component processes that run in parallel, 
the approach needs to allow variables local to a block 
to become shared variables of its component parallel processes.
To support the mechanisation of the rely/guarantee approach,
we have developed a synchronous concurrent refinement algebra.
Its foundation consists of a small set of primitive commands plus a small set of primitive operators
from which all remaining constructs are defined.
To support local variables we add a primitive \localisation\ operator to our algebra
that is used to define local variable blocks.
From this we can prove properties of localisation,
including its interaction with rely and guarantee conditions.
\end{abstract}

\ihin{R4 Poor introduction lacking motivation - what's the point? 

No connection to separation logic???}

\section{Introduction}\label{section:introduction}

Jones introduced a compositional approach to reasoning about concurrency
\cite{jones81d,Jones83a,Jones83b}
that makes use of rely and guarantee conditions.
To reason about an individual process, $P$, in isolation,
a rely condition, $r$, a binary relation on states, 
represents an assumption about the allowable interference the environment of $P$ 
can inflict on $P$ between the program steps of $P$.
Complementing this,
the processes in the environment of $P$ must guarantee 
that the steps they make satisfy a guarantee condition, also a binary relation on states,
such that the guarantee of each process implies the rely conditions of every process in its environment.
The semantics of rely/guarantee concurrency for shared memory
systems \cite{DaSMfaWSLwC} can be expressed using Aczel traces~\cite{Aczel83,DeRoever01}
that distinguish between steps of a process (program steps) and
steps of its environment (environment steps).

Our overall goal is to provide mechanised support for refining concurrent specifications 
in the rely/guarantee style to code.
Our approach has been to develop a core theory based on a few primitive commands
and a small set of operators.
The rely/guarantee concurrency theory is then built on top of that theory, 
and has been encoded in Isabelle/HOL \cite{Concurrent_Ref_Alg-AFP}.

The focus of this paper is on handling local variables in the context of concurrency.
In order to define a local variable block in terms of our core language, 
our semantics made use of a primitive ``\localisation'' operator \cite{DaSMfaWSLwC}.
The role of this paper is to explore the algebraic properties of the \localisation\ operator
in order to provide a basis for proving properties of concurrent programs involving local variables,
including proving properties about the interactions of local variables 
with rely and guarantee conditions within concurrent programs.
The \localisation\ operator has properties similar to 
those of existential quantification in predicate calculus $(\exists_x P)$,
for which Tarski developed cylindric algebra \cite{HenkinMonkTarski71}.
Predicate calculus can be used to represent properties of a single state,
whereas for concurrent programs one needs to consider properties of traces of states
and hence the localisation on concurrent programs, $\Hide{x}{c}$, 
localises a variable $x$ in every state of its traces.
Two additional variants are useful, 
one that localises $x$ in just the first state of its traces, $\HideFirst{x}{c}$,
and
one that localises $x$ in just the last state of its traces, $\HideLast{x}{c}$.

The approach needs to support nested local variable blocks,
including those for which an inner block reuses the same variable name as an outer block
(effectively masking the outer variable within the inner block).
This is an essential feature to support \ihsb{R3 more prominent}{(recursive)} procedures with local variables
and call-by-value parameters.

Sect.~\ref{section:CRA} gives an overview of our existing concurrent refinement algebra (CRA),
including our encoding of relies and guarantees within the core language.
Sect.~\ref{section:localisation} introduces \localisation\ by way of Tarski's cylindric algebra.
\Localisation\ over predicates is standard but in order to support localisation in CRA, 
we need to extend \localisation\ to binary relations and commands.
Sect.~\ref{section:commands} covers \localisation\ of commands, which includes
\localisation\ of tests via a straightforward lifting,
\localisation\ of atomic steps commands, again via lifting,
and
the interactions of \localisation\ with operators like sequential composition.
Sect.~\ref{section:iterations} proves some basic properties of \localisation s of iterations
and Sect.~\ref{section:rely-guar} uses these to prove properties of \localisation\ of
rely and guarantee commands.
Sect.~\ref{section:local-var} defines local variable blocks in terms of localisation and 
proves properties about how they interact with relies and guarantees.

\section{Concurrent refinement algebra}\label{section:CRA}

\ihsb{R4 this section done badly - source of issues}{}
To support rely/guarantee concurrency we have previously developed a \ihsb{R3 significance?}{synchronous}
concurrent refinement
algebra~\cite{FM2016atomicSteps,FMJournalAtomicSteps}.  The basis is a
complete distributive lattice $(\Commands, \leq, \Meet, \Join, \Magic, \Abort)$,
where the carrier set $\Commands$ can be interpreted as the set of
commands from a wide-spectrum language\footnote{A wide spectrum
  language includes both programming and specification constructs.};
meet, $\Meet$, as a strong form of command conjunction;
join, $\Join$, as nondeterministic choice;
least element $\Magic$ as the everywhere infeasible command (also known as magic);
and greatest element $\Abort$ as the everywhere aborting command (abort).
Binary meet and join operators are defined as 
$p \meet q \defs \Meet\{p,q\}$ and $p \join q \defs \Join\{p,q\}$.%
\footnote{Note that the lattice ordering $\leq$ is the reverse of the
  refinement ordering $\OLDrefsto$ that used in our earlier
  work~~\cite{FM2016atomicSteps,FMJournalAtomicSteps} and hence $\meet$ and $\join$ are also swapped, as are $\Magic$ and $\Abort$.}
\ihin{R3 explain why we have inverted lattice in footnote 2}

Two complete Boolean sub-algebras of the carrier set, $\Commands$, are
identified: one corresponding to the set of atomic program steps,
$\AtomicSteps$, and another to the set of instantaneous tests,
$\Tests$. The atomic steps and tests can be thought of as the primitive
commands from which others are generated through the application of
the program operators.

\subsubsection{The sub-algebra of tests.}

Letting $\Sigma \defs V \pfun S$ stand for the set of states that are
partial mappings from some countable set of variables $V$ to some set
$S$, we have that $(\pset(\Sigma), \subseteq$, $\Inter$, $\Union$,
$\bar{~~}$, $\emptyset$, $\Sigma)$ is a complete Boolean algebra of
predicates in which the order, $\subseteq$, is containment; meet,
$\Inter$, is intersection; join, $\Union$, is union; negation,
$\bar{~~~}$, is set complement; the least element is the empty set
$\emptyset$, and the greatest element is the universal set $\Sigma$.

The sub-algebra of tests, $\Tests$, is introduced to the refinement
algebra using an injective homomorphism, $\testsym: \pset(\Sigma) \fun
\Commands$, from the complete Boolean algebra of predicates to
commands.
In this way, test $\cgd{p}$ represents a command which terminates
instantaneously if the current state is in the set of states $p$, and
is otherwise infeasible. The meet of all tests, $\cgd{\emptyset}$, is
$\Magic$, the infeasible command, and the join of all tests,
$\cgd{\Sigma}$, is denoted by distinguished element $\Nil$,
which is the identity of sequential composition.

\subsubsection{The sub-algebras of atomic steps.}

For rely/guarantee reasoning, the carrier set of atomic steps,
$\AtomicSteps$, is further split into a complete Boolean
sub-algebra of program steps, $\PCommands$, and of environment steps
$\ECommands$.
The universe of binary relations on states,
$\pset{(\Sigma{\times}\Sigma)}$ also form a complete Boolean algebra,
and the sub-algebras of program and environments steps are introduced
using injective homomorphisms
$\programstep : \pset(\Sigma{\times}\Sigma) \fun \AtomicSteps$ and 
$\envstep : \pset(\Sigma{\times}\Sigma) \fun \AtomicSteps$.
Commands $\cpstep{r}$ and $\cestep{r}$ then represent commands that
can \ihsb{R1 do we really mean take?}{take} any atomic program, respectively environment, step satisfying
relation $r$.

Using $\universalrel \defs \Sigma \times \Sigma$ to represent the
universal relation on states, $\cpstepd \defs \pstep{\universalrel}$ is
the join of all program steps, and $\cestepd \defs
\cestep{\universalrel}$ is the join of all environment steps. Command
$\cpstepd$ is the complement of $\cestepd$ in the Boolean sub-algebra of atomic
steps, and we use 
$\cstepd  \defs  \cpstepd \join \cestepd$ 
to denote the join of all atomic steps. We also introduced the shorthand 
$\cstep{r} \defs \cpstep{r} \join \cestep{r}$ 
to represent any atomic step that updates the state of the program
according to relation $r$.  The meet of all program steps, all environment
steps, or all atomic steps are all $\Magic$, the infeasible
command which can take no atomic step at all.

\subsubsection{Sequential composition and iterations.}

The lattice is extended with a binary operator for sequential
composition (~$\Seq$~) that has identity $\Nil$. The fixed iteration
of a command $i\in \Nat$ times is denoted by $c^{i}$, and is defined
inductively by $c^{0} \defs \Nil$, $c^{i+1} \defs c \Seq
c^{i}$. Finite iteration ($\Fin{}$), possibly infinite iterations
($\Om{}$) and infinite iterations ($\Inf{}$) are defined using the
least ($\mu$) and greatest ($\nu$) fixed point operators of the
complete distributive lattice of commands.
They satisfy the isolation property (\refprop{isolation}) from \cite{Wright04}.
\vspace{-2ex}\\
\begin{minipage}[t]{0.5\textwidth}
\begin{eqnarray}
\Fin{c} &\defs& (\mu x \cdot \Nil \nondet c \Seq x) \labeldef{finite-iter}\\
\Om{c} &\defs& (\nu x \cdot \Nil \nondet c \Seq x) \labeldef{omega-iter}
\end{eqnarray}
\end{minipage}%
\begin{minipage}[t]{0.5\textwidth}
\begin{eqnarray}
\Inf{c} &\defs& \Om{c} \Seq \Magic \labeldef{inf-iter-def}\\
\Om{c} & = & \Fin{c} \nondet \Inf{c} \labelprop{isolation}
\end{eqnarray}
\end{minipage}

\subsubsection{Synchronous parallel and weak conjunction operators.}

Two binary synchronous operators are \ihsb{R3 introduced to what? Types?}{introduced}: one for parallel
composition ($\parallel$) and one for weak conjunction $(\together)$,
where the behaviour of weak conjunction coincides with the lattice
meet up until failure of either command, at which point it behaves
like the aborting command $\Abort$.

The identity of parallel composition is the command $\Skip \defs
\Om{\cestepd}$, the possibly infinite iteration of environment steps,
and the identity of weak conjunction is the command $\Chaos \defs \Om{\cstepd}$, the
possibly infinite iteration of any atomic step.

\subsubsection{Rely, guarantee and demand commands.}

The command $(\guar{g})$ is the least command refining $\Chaos$ that
\ihsb{R3 what does constrain mean?}{constrains} every program step taken to satisfy relation
$g$. Similarly, command $(\Demand{r})$ constrains environment steps to
satisfy relation $r$.
The command $(\rely{r})$ does not constrain the atomic steps taken by
the command in any way, but can abort if an environment step is taken
that does not satisfy $r$:\vspace{-3ex}\\
\begin{minipage}[t]{0.5\textwidth}
\begin{eqnarray}
  \guar{g} & \defs &   \Om{(\cpstep{g} \nondet \cestepd)} \labeldef{guar} \\
  \Demand{r} & \defs & \Om{(\cestep{r} \nondet \cpstepd)} \labeldef{demand} 
\end{eqnarray}
\end{minipage}%
\begin{minipage}[t]{0.5\textwidth}
\begin{eqnarray}
  \rely{r} & \defs &   \Om{(\cstepd \nondet \cestep{\overline{r}} \Seq \Abort)} \labeldef{rely}\\
  \rely{r} & = & \Om{\cstepd} \Seq (\Nil \nondet \cestep{\overline{r}} \Seq \Abort) \labelprop{rely-rewritten}
\end{eqnarray}
\end{minipage}\vspace{1ex}
Using the \ihsb{R4 not intuitive}{decomposition rule}, $\Om{(c \nondet d)} = \Om{c} \Seq \Om{(d \Seq \Om{c})}$,
and simplifying we have that rely commands satisfy equivalence (\refprop{rely-rewritten}).
Command $(\guar{\id})$, where $\id$ is the identity relation,
describes a command which can only take \ihsb{R4 not defined}{stuttering} program
steps. Combining it with $\Term$, the command that terminates if it is
not interrupted by its environment forever, defines the command
$\Idle$.\vspace{-4ex}\\
\begin{minipage}[t]{0.5\textwidth}
\begin{eqnarray}
\Term & \defs & \Fin{\cstepd} \Seq \Om{\cestepd} \labeldef{term}
\end{eqnarray}
\end{minipage}%
\begin{minipage}[t]{0.5\textwidth}
\begin{eqnarray}
\Idle & \defs & \Term \together (\guar{\id}) \labeldef{idle}
\end{eqnarray}
\end{minipage}\vspace{1ex}
These commands satisfy the following properties \cite{FM2016atomicSteps,FMJournalAtomicSteps}.
\ihsb{R4 no motivation for these properties}{}
\begin{eqnarray}
  (\guar{g_1}) \together (\guar{g_2}) & = & \guar{(g_1 \inter g_2)} \labelprop{merge-guarantees} \\
  (\guar{g}) \together (c \Seq d) & = & ((\guar{g}) \together c) \Seq ((\guar{g}) \together d) \labelprop{guar-distrib-seq} \\
  (\rely{r_1}) & \refsto & (\rely{r_2}) ~~~~~~~~~~~~~~~~~~~\mbox{if $r_1 \subseteq r_2$} \labelprop{weaken-rely} \\
  (\rely{r}) \together c & \refsto & c \labelprop{remove-rely} \\
  (\rely{r_1}) \together (\rely{r_2}) & = & \rely{(r_1 \inter r_2)} \labelprop{merge-relies} \\
  (\rely{r}) \together (c \Seq d) & \refsto & ((\rely{r}) \together c) \Seq ((\rely{r}) \together d) \labelprop{rely-distrib-seq} \\
  (\Demand{r}) \together (\rely{r}) & = & (\Demand{r})  \labelprop{demand-rely} \\
  \Idle & = & \Idle \together (\guar{g})  \mbox{~~~~~~if $g$ is reflexive} \labelprop{idle-together-reflexive-guarantee}
\end{eqnarray}

\vspace{-3ex}
\section{Cylindric algebras and localisation}\label{section:localisation}

Localisation (or hiding) of variables in programs has similarities to
quantification in predicate calculus.  In the same way that $\exists_{x} p$
localises $x$ within the quantification, one can define an
operator $\Hide{x}{c}$ that localises variable $x$ within the command
$c$.
Some key algebraic similarities between these operators are abstracted by
Tarski's cylindric algebras~\cite{HenkinMonkTarski71}, which are used
in algebraic logic for first order predicate calculus.
Before extending the refinement algebra to include a notion of
localisation, we introduce two fundamental cylindric algebras, 
and show how they can be instantiated for Boolean
algebras of predicates and relations.
A complete cylindric algebra (without diagonals) extends a complete
lattice with a cylindrical operator, $\Hide{x}{}$, for each variable $x$.

\begin{definitionx}[complete cylindric algebra]
\ihsb{R3 who's definition? HMT or ours?}{}
A complete cylindric algebra is a complete lattice $(L, \leq,
\bigsqcup, \bigsqcap, \bot, \top)$, extended with a countable set of
variables $V$ and a new unary operator $\Hide{x}{}$ on $L$ defined for
each $x\in V$. The new operator satisfies the following axioms in
which $p \in L$ and $P \subseteq L$.\vspace{-3ex}\\
\begin{minipage}[t]{0.5\textwidth}
\begin{eqnarray}
p & \leq & \Hide{x}{p}      \labelax{hide-weakens} \\
\Hide{x}{(p \sqcap (\Hide{x}{q}))} & = & (\Hide{x}{p}) \sqcap (\Hide{x}{q})  \labelax{hide-distrib-meet} 
\end{eqnarray}
\end{minipage}%
\begin{minipage}[t]{0.5\textwidth}
\begin{eqnarray}
\Hide{x}{\Hide{y}{p}} & = & \Hide{y}{\Hide{x}{p}}  \labelax{hide-commutes}\\
\Hide{x}{(\bigsqcup P)} & = & \bigsqcup_{p \in P} (\Hide{x}{p})   \labelax{hide-distrib-Join}
\end{eqnarray}
\end{minipage}
\end{definitionx}
From these axioms one can derive the following basic properties.\vspace{-3ex}\\
\begin{minipage}[t]{0.5\textwidth}
\begin{eqnarray}
  \Hide{x}{\bot} & = & \bot  \labelprop{hide-magic} \\
  \Hide{x}{\top} & = & \top   \labelprop{hide-abort} \\
  \Hide{x}{(p \sqcup q)} & = & (\Hide{x}{p}) \sqcup (\Hide{x}{q})   \labelprop{hide-distrib-binary-join}
\end{eqnarray}
\end{minipage}%
\begin{minipage}[t]{0.5\textwidth}
\begin{eqnarray}
  \Hide{x}{\Hide{x}{p}} & = & \Hide{x}{p}  \labelprop{hide-idempotent} \\
  \Hide{x}{p} & \leq & \Hide{x}{q} ~~~~~~~\mbox{if $p \leq q$}  \labelprop{hide-mono}
\end{eqnarray}
\end{minipage}\vspace{1ex}
Property (\refprop{hide-magic}) follows from (\refax{hide-distrib-Join}) by taking $P$ to be the empty set.
Property (\refprop{hide-abort}) follows from (\refax{hide-weakens}) and the fact that $\top$ is the greatest element.
Property (\refprop{hide-distrib-binary-join}) follows as the binary case of (\refax{hide-distrib-Join}).
Property (\refprop{hide-idempotent}) follows from (\refax{hide-distrib-meet}) taking $p$ to be $\Abort$ and then applying (\refprop{hide-abort}).
Property (\refprop{hide-mono}) follows from (\refprop{hide-distrib-binary-join}) and
the property of lattices that $p \leq q$ if $p \sqcup q = q$.
It is also useful to consider the case where the complete lattice also
contains a negation operator, and forms a Boolean algebra.

\begin{definitionx}[complete Boolean cylindric algebra]
A complete Boolean cylindric algebra is a complete Boolean algebra
$(L, \leq, \bigsqcup, \bigsqcap, \bar{~~}, \bot, \top)$, with negation
operator $\bar{~~}$, extended with a countable set of variables $V$ and a
new unary operator $\Hide{x}{}$ on $L$ defined for each $x\in V$,
satisfying axioms 
(\refax{hide-weakens}--\refax{hide-distrib-Join}).
\end{definitionx}

In a complete Boolean cylindric algebra, the dual of $\Hide{x}{p}$ is
defined as follows.
\begin{eqnarray}
  \HideUniv{x}{p} & \defs & \Complement{\Hide{x}{\Complement{p}}} \labeldef{hide-univ}
\end{eqnarray}
Given an intuitive interpretation of $\Hide{x}{p}$ as existential
quantification, its dual, $\HideUniv{x}{p}$, can be thought of as
universal quantification. It satisfies, for example,
\begin{eqnarray}
\Hide{x}{\HideUniv{x}{p}} & = & \HideUniv{x}{p}~. \labelprop{localise-universal} 
\end{eqnarray}

\subsubsection{Localisations over predicates.}

Given a representation of predicates as \ihsb{Meaning? power set?}{the universe of a state space}
$\Sigma \defs V \pfun S$, where $V$ is a countable set of variables,
and $S$ is some set of possible values for those variables, 
a cylindrification operator for the complete Boolean algebra of
predicates
is 
\begin{eqnarray}
\Hide{x}{p} & \defs & \id(\overline{x}) \limg p \rimg \labeldef{hide-predicate}
\end{eqnarray}
where for a set of variables $vs$,
$\id(vs)$ is the identity relation on $V$
(i.e.\ variables in $vs$ map to themselves and all other variables are unconstrained),
$\overline{x}$ is the set of variables other than $x$,
and $r \limg p \rimg$ is the image of the set $p$ through the relation $r$.

\subsubsection{Localisations over relations.}

For the complete Boolean algebras of binary relations on state space
$\Sigma \defs V \pfun S$ 
with binary \ihsb{R3 relation to thin ;}{relational composition} \ihsb{R4 messay}{represented by} ``$\semi$'', 
one can distinguish three interesting cylindrifications:\vspace{-3ex}\\
\begin{minipage}[t]{0.5\textwidth}
\begin{eqnarray}
\HideFirst{x}{r} & \defs & (\id(\overline{x}) \semi r) \labeldef{hide-first-relation} \\
\HideLast{x}{r}  & \defs & (r \semi \id(\overline{x})) \labeldef{hide-last-relation}
\end{eqnarray}
\end{minipage}%
\begin{minipage}[t]{0.5\textwidth}
\begin{eqnarray}
\Hide{x}{r}      & \defs & (\id(\overline{x}) \semi r \semi \id(\overline{x}))  \labeldef{hide-relation} 
\end{eqnarray}
\end{minipage}\vspace{1ex}
where the first localises variable $x$ from $V$ in the initial state of the relation $r$;
the second localises $x$ in the final state of the relation, and
the last localises $x$ in both the first and last states.
For these instantiations, all of these cylindrifications can be shown
to commute, e.g. $\HideFirst{x}{\HideLast{y}{r}} =
\HideLast{y}{\HideFirst{x}{r}}$, and satisfy the following relationships:\vspace{-3ex}\\
\begin{minipage}[t]{0.5\textwidth}
\begin{eqnarray}
\HideFirst{x}{r} & \subseteq & \Hide{x}{r}  \labelprop{hide-first-both-rel}
\end{eqnarray}
\end{minipage}
\begin{minipage}[t]{0.5\textwidth}
\begin{eqnarray}
\HideLast{x}{r} & \subseteq & \Hide{x}{r}  \labelprop{hide-last-both-rel}
\end{eqnarray}
\end{minipage}\vspace{1ex}
Additionally, hiding a variable $x$ in both the first and last state
of a relation is equivalent to hiding the variable in the entire
relation. 
\begin{eqnarray}
\Hide{x}{r} & = & \HideFirst{x}{\HideLast{x}{r}} = \HideLast{x}{\HideFirst{x}{r}} \labelprop{hide-both-rel}
\end{eqnarray}
Note that this last property does not extend to arbitrary commands (see the next section).

\section{Localisations over commands}\label{section:commands}

To reason about localisation of variables in commands we extend the
complete distributive lattice of the concurrent refinement algebra with a
further with three additional cylindrifications, $\HideFirst{x}{}$,
$\HideLast{x}{}$ and $\Hide{x}{}$, defined for each variable $x$ from
a countable set $V$ of variables. These have an interpretation over
commands as operators that localise $x$ in the first state of the
command, the last state of the command, and in \ihsb{R3 when did they become traces?}{all states of the
command}, respectively. Together they are assumed to satisfy the axioms\vspace{-3ex}\\
\begin{minipage}[t]{0.4\textwidth}
\begin{eqnarray}
\HideFirst{x}{c} & \leq & \Hide{x}{c}  \labelax{hide-first-both} \\
\HideLast{x}{c} & \leq & \Hide{x}{c}  \labelax{hide-last-both}
\end{eqnarray}
\end{minipage}%
\begin{minipage}[t]{0.6\textwidth}
\begin{eqnarray}
\Hide{x}{\HideFirst{y}{c}}     & = & \HideFirst{y}{\Hide{x}{c}} \labelax{hide-first-commutes} \\
\Hide{x}{\HideLast{y}{c}}      & = & \HideLast{y}{\Hide{x}{c}} \labelax{hide-last-commutes} \\
\HideFirst{x}{\HideLast{y}{c}} & = & \HideFirst{y}{\HideLast{x}{c}} \labelax{first-last-commutes}
\end{eqnarray}
\end{minipage}\vspace*{2ex}\\
with the first two corresponding to the intuition that localising
variable $x$ in either the first or last state of a command is less
non-deterministic than localising $x$ in all states of the command
(i.e.~first, last and all intermediate states). The last three describe
commutativity of the cylindrifications.
From these axioms and the axioms of the cylindrification operators we
can prove, for example, that localising $x$ in the first or last state
of a command, has no effect on a command in which $x$ has already been
localised in all states:
\begin{minipage}{0.5\textwidth}
\begin{eqnarray}
\Hide{x}{c} & = & \HideFirst{x}{(\Hide{x}{c})} = \Hide{x}{(\HideFirst{x}{c})} \labelprop{hide-all-then-first}
\end{eqnarray}
\end{minipage}%
\begin{minipage}{0.5\textwidth}
\begin{eqnarray}
\Hide{x}{c} & = & \HideLast{x}{(\Hide{x}{c})} = \Hide{x}{(\HideLast{x}{c})} \labelprop{hide-all-then-last}
\end{eqnarray}
\end{minipage}\\ \\
We elide an axiom like (\refprop{hide-both-rel}) because commands may
take an arbitrary number of steps, and so localising the first and
last state only may not be sufficient to localise all of the
intermediate states.

\subsubsection{The sub-algebra of tests.}

For our cylindrifications we require that the mapping $\testsym$ from
predicates to test commands preserves the $\Hide{x}{}$
cylindrification operator:
\begin{eqnarray}
\Hide{x}{\cgd{p}} & = & \cgd{\Hide{x}{p}}    \labelax{hide-test}
\end{eqnarray}
and we define all three cylindrifications to have the same meaning on tests: 
\begin{eqnarray}
\Hide{x}{\cgd{p}} & = &
\HideFirst{x}{\cgd{p}} =
\HideLast{x}{\cgd{p}}
\labelax{hide-test-equality}
\end{eqnarray}
Using (\refprop{hide-abort}) we can show that localisation of any
variable $x$ in $\Nil = \cgd{\Sigma}$ has no effect.
\begin{eqnarray}
\Nil & = &
\HideFirst{x}{\Nil} =
\HideLast{x}{\Nil} = 
\Hide{x}{\Nil} 
\labelprop{hide-nil}
\end{eqnarray}

\subsubsection{The sub-algebras of program and environment steps.}

For program and environment steps we require the mappings
$\programstep$ and $\envstep$ to preserve all three cylindrification
operators: \vspace{-2ex}\\
\begin{minipage}{0.5\textwidth}
\begin{eqnarray}
  \HideFirst{x}{\cpstep{r}} & = & \cpstep{\HideFirst{x}{r}}  \labelax{hide-pgm-first} \\ 
  \HideLast{x}{\cpstep{r}} & = & \cpstep{\HideLast{x}{r}}  \labelax{hide-pgm-last} \\ 
  \Hide{x}{\cpstep{r}} & = & \cpstep{\Hide{x}{r}}  \labelax{hide-pgm}
\end{eqnarray}
\end{minipage}%
\begin{minipage}{0.5\textwidth}
\begin{eqnarray}
  \HideFirst{x}{\cestep{r}} & = & \cestep{\HideFirst{x}{r}}  \labelax{hide-env-first} \\ 
  \HideLast{x}{\cestep{r}} & = & \cestep{\HideLast{x}{r}}  \labelax{hide-env-last} \\ 
  \Hide{x}{\cestep{r}} & = & \cestep{\Hide{x}{r}}  \labelax{hide-env}
\end{eqnarray}
\end{minipage}\vspace{1ex}
from which, using (\refprop{hide-both-rel}), we can show that for any
relation $r$,
\begin{eqnarray}
\Hide{x}{\cpstep{r}} & = &
\HideFirst{x}{\HideLast{x}{\cpstep{r}}} = \HideLast{x}{\HideFirst{x}{\cpstep{r}}}
\labelprop{hide-both-prog}\\
\Hide{x}{\cestep{r}} & = &
\HideFirst{x}{\HideLast{x}{\cestep{r}}} = \HideLast{x}{\HideFirst{x}{\cestep{r}}}
\labelprop{hide-both-env}\\
\Hide{x}{\cstep{r}} & = &
\HideFirst{x}{\HideLast{x}{\cestep{r}}} = \HideLast{x}{\HideFirst{x}{\cstep{r}}}
\labelprop{hide-both-atomic}
\end{eqnarray}
and that localisation of any variable $x$ in $\cpstepd$, $\cestepd$ or $\cstepd$
has no effect:\vspace{-2ex}\\
\begin{minipage}[t]{0.5\textwidth}
\begin{eqnarray}
\cpstepd & = &
\HideFirst{x}{\cpstepd} =
\HideLast{x}{\cpstepd} = 
\Hide{x}{\cpstepd} 
\labelprop{hide-cpstepd} \\
\cestepd & = &
\HideFirst{x}{\cestepd} =
\HideLast{x}{\cestepd} = 
\Hide{x}{\cestepd} 
\labelprop{hide-cestepd}
\end{eqnarray}
\end{minipage}%
\begin{minipage}[t]{0.5\textwidth}
\begin{eqnarray}
\cstepd & = &
\HideFirst{x}{\cstepd} =
\HideLast{x}{\cstepd} = 
\Hide{x}{\cstepd} 
\labelprop{hide-cstepd}
\end{eqnarray}
\end{minipage}

\subsubsection{Sequential composition.}

Because sequential composition is not commutative,
defining localisation over a sequential composition requires two
axioms similar to (\refax{hide-distrib-meet}). \vspace{-2ex}\\
\begin{minipage}{0.5\textwidth}
\begin{eqnarray}
  \Hide{x}{(c \Seq (\HideFirst{x}{d}))} & = & (\Hide{x}{c}) \Seq (\Hide{x}{d}) \labelax{hide-seq-right-strong}
\end{eqnarray}
\end{minipage}%
\begin{minipage}{0.5\textwidth}
\begin{eqnarray}
  \Hide{x}{((\HideLast{x}{c}) \Seq d)} & = & (\Hide{x}{c}) \Seq (\Hide{x}{d}) \labelax{hide-seq-left-strong}
\end{eqnarray}
\end{minipage}\vspace{1ex}
From these one can derive the following using (\refax{hide-first-both}), (\refax{hide-last-both}) and (\refprop{hide-idempotent}).\vspace{-2ex}\\
\begin{minipage}{0.5\textwidth}
\begin{eqnarray}
  \Hide{x}{(c \Seq (\Hide{x}{d}))} & = & (\Hide{x}{c}) \Seq (\Hide{x}{d}) \labelprop{hide-seq-right}
\end{eqnarray}
\end{minipage}%
\begin{minipage}{0.5\textwidth}
\begin{eqnarray}
  \Hide{x}{((\Hide{x}{c}) \Seq d)} & = & (\Hide{x}{c}) \Seq (\Hide{x}{d}) \labelprop{hide-seq-left}
\end{eqnarray}
\end{minipage}\vspace{1ex}

We say that a command $c$ takes at least one atomic step when $c = c
\together (\alpha\Seq\Chaos)$. For such a command $c$, and arbitrary
command $d$, we introduce the following axioms:
\begin{eqnarray}
\HideFirst{x}{(c \Seq d)} & = & (\HideFirst{x}{c}) \Seq d
\textrm{~~~~if~$c = c \together (\alpha \Seq \Chaos)$}
\labelax{hide-first-seq-atomic} \\
\HideLast{x}{(d \Seq c)} & = & d \Seq (\HideLast{x}{c})
\textrm{~~~~if~$c = c \together (\alpha \Seq \Chaos)$}
\labelax{hide-last-seq-atomic}
\end{eqnarray}
Note that these do not hold if $c$ is a test, which is instantaneous
(i.e.\ it does not take any atomic steps). 

\subsubsection{Synchronization operators.}

For synchronization operator $\sync$ taken to be either parallel
($\parallel$) or weak conjunction $(\together)$, we have that our
cylindrifications are assumed to satisfy the following axioms 
similar to (\refax{hide-distrib-meet}).
\begin{eqnarray}
\Hide{x}{(c \sync (\Hide{x}{d}))} & = & 
(\Hide{x}{c}) \sync (\Hide{x}{d}) 
\labelax{hide-distrib-sync} \\
\HideFirst{x}{(c \sync (\HideFirst{x}{d}))} & = &
(\HideFirst{x}{c}) \sync (\HideFirst{x}{d})   
\labelax{hide-first-distrib-sync}
\end{eqnarray}
We intentionally elide the corresponding property for the
$\HideLast{x}{}$ cylindrical operator, because it does not hold in our
intended semantic model -- take for instance a scenario where $c$
and $d$ take different numbers of atomic steps.

\section{Localisations of iterations}\label{section:iterations}

\newcommand{\Fixed}[2]{#1^{#2}}

An iteration of a \localised\ command is \localised. 

\begin{lemmax}[localised-fixed-iteration]
\begin{math}
\Hide{x}{\Fixed{(\Hide{x}{c})}{i}} = \Fixed{(\Hide{x}{c})}{i}
\end{math}
\end{lemmax}

\begin{proof}
The proof is by induction on $i$.
The base case follows from $\Fixed{c}{0} = \Nil$ and (\refprop{hide-nil}).
The inductive case unfolds the iteration once 
(i.e.~$ \Fixed{(\Hide{x}{c})}{i+1} =  (\Hide{x}{c}) \Seq\Fixed{(\Hide{x}{c})}{i}$), 
applies (\refprop{hide-seq-left}) followed by the induction hypothesis
and then refolds the iteration.
\end{proof}

\begin{lemmax}[localised-finite-iteration]
\begin{math}
\Hide{x}{\Fin{(\Hide{x}{c})}} = \Fin{(\Hide{x}{c})}.
\end{math}
\end{lemmax}

\begin{proof}
The proof uses the fact that $\Fin{c} = \Join_{i \in \nat} \Fixed{c}{i}$,
applies (\refax{hide-distrib-Join}) and then \reflem{localised-fixed-iteration}.
\end{proof}
  
\begin{lemmax}[localised-omega-iteration]
\begin{math}
\Hide{x}{\Om{(\Hide{x}{c})}} = \Om{(\Hide{x}{c})}.
\end{math}
\end{lemmax}

\begin{proof}
From (\refax{hide-weakens}) it is enough to show
$\Om{(\Hide{x}{c})} \refsto \Hide{x}{\Om{(\Hide{x}{c})}}$.
Using fixed-point induction \cite{fixedpointcalculus1995} 
based on the definition of possibly infinite iteration (\refdef{omega-iter})
it suffices to prove
\begin{displaymath}
(\Hide{x}{c}) \Seq (\Hide{x}{\Om{(\Hide{x}{c})}}) \nondet \Nil \refsto \Hide{x}{\Om{(\Hide{x}{c})}} ~.
\end{displaymath}
Starting from the right side, the proof is as follows.
\begin{displaymath}
\Hide{x}{\Om{(\Hide{x}{c})}}
\Equals*[unfold iteration]
\Hide{x}{((\Hide{x}{c}) \Seq \Om{(\Hide{x}{c})} \nondet \Nil)}
\Equals*[distribute \localisation\ over binary choice (\refprop{hide-distrib-binary-join})]
\Hide{x}{((\Hide{x}{c}) \Seq \Om{(\Hide{x}{c})})} \nondet (\Hide{x}{\Nil})
\Equals*[using (\refprop{hide-seq-left}) and (\refprop{hide-nil})]
(\Hide{x}{c}) \Seq (\Hide{x}{\Om{(\Hide{x}{c}})}) \nondet \Nil
\end{displaymath}
\end{proof}

\begin{lemmax}[localised-atomic-fixed-iteration]
For atomic step $a$ and variable $x$, 
if $(\Hide{x}{a}) = (\HideLast{x}{a})$, then
\(
\Hide{x}{(\Fixed{a}{i})} = \Fixed{(\Hide{x}{a})}{i}.
\)
\end{lemmax}

\begin{proof}
The proof is by induction on $i$.
The base case uses the property that $\Fixed{c}{0} = \Nil$ and (\refprop{hide-nil}).
For the inductive case we assume that our property holds for $i$ and
show that it holds for $i+1$:
\begin{displaymath}
\Hide{x}{(\Fixed{a}{i+1})}
\Equals*[unfold fixed iteration and (\refprop{hide-all-then-last})]
\Hide{x}{(\HideLast{x}{(\Fixed{a}{i} \Seq a)})}
\Equals*[by (\refax{hide-last-seq-atomic}) as $a = a \together \cstepd \Seq \Chaos$]
\Hide{x}{(\Fixed{a}{i} \Seq (\HideLast{x}{a}))}
\Equals*[assumption $(\Hide{x}{a}) = (\HideLast{x}{a})$ and (\refprop{hide-seq-right})]
(\Hide{x}{(\Fixed{a}{i})}) \Seq (\Hide{x}{a})

\Equals*[inductive assumption and fold fixed iteration]
\Fixed{(\Hide{x}{a})}{i+1}
\end{displaymath}
\end{proof}

\begin{lemmax}[localised-atomic-finite-iteration]
For atomic step $a$ and variable $x$,
if $(\Hide{x}{a}) = (\HideLast{x}{a})$, then
\(
  \Hide{x}{(\Fin{a})} = \Fin{(\Hide{x}{a})}~.
\)
\end{lemmax}

\begin{proof}
The proof follows from the fact that $\Fin{c} = \Join_{i \in \nat} \Fixed{c}{i}$,
and applies (\refax{hide-distrib-Join}) followed by \reflem{localised-atomic-fixed-iteration}
using the assumption.
\end{proof}

In order show the equivalent of
\reflem{localised-atomic-finite-iteration} for possibly infinite
iteration, we need to assume an additional axiom about infinite
iteration
\begin{eqnarray}
\Hide{x}{(\Inf{a})} & = & \Inf{(\Hide{x}{a})}
\textrm{~~~~~ if $(\Hide{x}{a}) = (\HideLast{x}{a})$}
\labelax{hide-atomic-infinite-iteration}
\end{eqnarray}
which holds in our intended semantic model~\cite{DaSMfaWSLwC}. It is
believed (but not proven) to be independent of the other
axioms.\footnote{For instance it is possible to verify that
  assumption $(\Hide{x}{a}) = (\HideLast{x}{a})$ implies
  $\Fin{(\Hide{x}{a})}\Seq \Magic \leq \Hide{x}{(\Inf{a})}$ using the
  current axiomatisation. However, extending that reasoning to the
  infinite case does not seem to be possible without further
  continuity assumptions.}

\begin{lemmax}[localised-atomic-omega-iteration]
For atomic step $a$ and variable $x$, assuming
(\refax{hide-atomic-infinite-iteration}) and $(\Hide{x}{a}) =
(\HideLast{x}{a})$, then \( \Hide{x}{(\Om{a})} = \Om{(\Hide{x}{a})}~.
\)
\end{lemmax}

\begin{proof}
The proof follows from isolation (\refprop{isolation}),
property (\refprop{hide-distrib-binary-join}), and
\reflem{localised-atomic-finite-iteration} and
(\refax{hide-atomic-infinite-iteration}) using assumption
$(\Hide{x}{a}) = (\HideLast{x}{a})$.
\end{proof}

\section{Localisations over rely/guarantee constructs}\label{section:rely-guar}

Localisation of a rely condition can be simplified in the following way.
\begin{lemmax}[localise-rely]
For relation $r$ and variable $x$, 
\(
  \Hide{x}{(\rely{r})} = \rely{(\HideUniv{x}{r})}~.
\)
\end{lemmax}
\begin{proof}
\begin{displaymath}
\Hide{x}{(\rely{r})}
\Equals*[rewrite rely using (\refprop{rely-rewritten})]
\Hide{x}{(\Om{\cstepd} \Seq (\Nil \nondet \cestep{\overline{r}}\Seq \Abort))}
\Equals*[$\Om{\cstepd} = \Hide{x}{(\Om{\cstepd})}$ by
  (\refprop{hide-cstepd}) and \reflem{localised-omega-iteration}, 
  and (\refprop{hide-seq-left})]
\Om{\cstepd} \Seq (\Hide{x}{(\Nil \nondet \cestep{\overline{r}}\Seq \Abort)})
\Equals*[by distributivity of \localisation\ over binary choice (\refprop{hide-distrib-binary-join})]
\Om{\cstepd} \Seq ((\Hide{x}{\Nil}) \nondet (\Hide{x}{(\cestep{\overline{r}} \Seq \Abort)}))
\Equals*[by (\refprop{hide-nil}), property (\refprop{hide-abort}) and (\refprop{hide-seq-right})]
\Om{\cstepd} \Seq (\Nil \nondet (\Hide{x}{\cestep{\overline{r}}}) \Seq \Abort)
\Equals*[by (\refax{hide-env})]
\Om{\cstepd} \Seq (\Nil \nondet \cestep{\Hide{x}{\overline{r}}} \Seq \Abort)
\Equals*[rewrite rely using (\refprop{rely-rewritten}) and (\refdef{hide-univ})]
\rely{(\HideUniv{x}{r})}
\end{displaymath}    
\end{proof}

From \reflem{localise-rely} we have that $\rely{(\Hide{x}{r})} =
\Hide{x}{(\rely{r})}$ if $r = \Hide{x}{r}$. For guarantee and demand
conditions, a weaker condition suffices.
\begin{lemmax}[localise-gdr]
For relation $r$, and variable $x$ we have that
\ihsb{R3 discuss assumptions on these}{}
\begin{eqnarray}
  \guar{(\Hide{x}{r})} & = & \Hide{x}{(\guar{r})}
  \textrm{~~~~~~~~~~if $(\Hide{x}{r}) = (\HideLast{x}{r})$}
  \labelprop{hide-guar}\\
  \Demand{(\Hide{x}{r})} & = & \Hide{x}{(\Demand{r})}
  \textrm{~~~~if $(\Hide{x}{r}) = (\HideLast{x}{r})$}
  \labelprop{hide-demand} \\
  \rely{(\Hide{x}{r})} & = & \Hide{x}{(\rely{r})}
  \textrm{~~~~~~~~~~~if $(\Hide{x}{r}) = r$}
  \labelprop{hide-rely}
\end{eqnarray}
\end{lemmax}

\begin{proof}
From (\reflem{localised-atomic-omega-iteration}) and the guarantee
definition (\refdef{guar}) it is enough to show for
(\refprop{hide-guar}) that:
\begin{displaymath}
\Hide{x}{(\cestepd \nondet \cpstep{r})}
\Equals*[by distributivity over binary choice (\refprop{hide-distrib-binary-join})]
(\Hide{x}{\cestepd}) \nondet (\Hide{x}{\cpstep{r}})
\Equals*[by (\refprop{hide-cestepd}), (\refax{hide-pgm}), assumption $(\Hide{x}{r}) = (\HideLast{x}{r})$ and (\refax{hide-pgm-last})]
(\HideLast{x}{\cestepd}) \nondet (\HideLast{x}{\cpstep{r}}) 
\Equals*[by distributivity over binary choice (\refprop{hide-distrib-binary-join})]
\HideLast{x}{(\cestepd \nondet \cpstep{r})}
\end{displaymath}
and the proof for (\refprop{hide-demand}) is similar.
Property (\refprop{hide-rely}) follows from \reflem{localise-rely} and (\refprop{localise-universal}). 
\end{proof}
From \reflem{localise-gdr} we have, for example, that the following
hold because $(\HideLast{x}{\id(x)}) = (\Hide{x}{\id(x)}) = \universalrel$.
\begin{eqnarray}
\Hide{x}{(\guar{(\id(x))})} & = & \guar{(\Hide{x}{\id(x)})} = \guar{(\universalrel)} = \Chaos \labelprop{hide-guar-id-x} \\
\Hide{x}{(\demand{\id(x)})} & = & \Demand{(\Hide{x}{\id(x)})} = \Demand{(\universalrel)} = \Chaos \labelprop{hide-demand-id-x}
\end{eqnarray}

\section{Local variable blocks in concurrent refinement algebra}\label{section:local-var}

Local variable blocks are defined in terms of a construct that
introduces a local scope $\Local{x}{c}$ for a variable $x$ around a
command $c$.
Within the local scope, $\Local{x}{c}$, the local
variable $x$ is not modified by its environment.\footnote{The command
  $\Demand{\id(x)}$ restricts all environment steps to not modify $x$
  but puts no constraints on program steps (\refdef{demand}).}
From outside of the local scope of $\Local{x}{c}$, non-local $x$ is
guaranteed not to by modified by the program steps of the command, and
is unconstrained by its environment steps. This behaviour is obtained
by \localising\ $x$ in $(c \together \Demand{\id{(x)}})$, which can be
thought of as obscuring the local behaviour of variable $x$, and then
conjoining the result with $\guar{\id(x)}$ to give (\refdef{localise}).

A local variable block is then defined in terms of localisation by
adding $\Idle$ commands (\refdef{idle}) to allow stuttering program
steps that may be required to allocate and deallocate the local
variable (\refdef{var-intro}).  The following definitions are as in
\cite{DaSMfaWSLwC}.
\begin{eqnarray}
\Local{x}{c} & \defs & \Hide{x}{(\eguard{\id(x)} \together c)} \together \guar{\id(x)} \labeldef{localise}\\
\variable{x}{c} & \defs & \Idle \Seq \Local{x}{c} \Seq \Idle \labeldef{var-intro}
\end{eqnarray}

To prove properties of local variable blocks, equivalent properties
are first proven for localisations.
Directly from the definition of (\refdef{localise}), we have that
local variable block $\Local{x}{c}$ guarantees $\id(x)$ for non-local
$x$.
\begin{lemmax}[local-default-guarantee]
\(
  \Local{x}{c} = \Local{x}{c} \together \guar{\id(x)}
\)
\end{lemmax}

\begin{proof}
By definition (\refdef{localise}) and $\together$ is idempotent.
\end{proof}

By definition (\refdef{idle}), the command $\Idle$ satisfies the
guarantee $\id(x)$, and so this result can be promoted to local
variable blocks.
\begin{lemmax}[variable-default-guarantee]
\(
  \variable{x}{c} = \variable{x}{c} \together \guar{\id(x)}
\)
\end{lemmax}

\begin{proof}
\begin{displaymath}
\variable{x}{c}
\Equals*[expand local variable block definition (\refdef{var-intro})]
\Idle \Seq \Local{x}{c} \Seq \Idle 
\Equals*[by \reflem{local-default-guarantee}, definition of $\Idle$ (\refdef{idle}) and (\refprop{merge-guarantees})]
(\Idle \together \guar{\id(x)}) \Seq
(\Local{x}{c} \together \guar{\id(x)}) \Seq
(\Idle \together \guar{\id(x)})
\Equals*[distribute guarantee over sequential composition by (\refprop{guar-distrib-seq})]
(\Idle \Seq
\Local{x}{c} \Seq
\Idle) \together \guar{\id(x)}
\Equals*[fold local variable block definition (\refdef{var-intro})]
\variable{x}{c} \together \guar{\id(x)}
\end{displaymath}
\end{proof}

For any local scope $\variable{x}{c}$, there is an implicit
$\demand{\id(x)}$ within the local scope, and so we can rely on the
environment to not modify local $x$ within that scope.
This result is trivially promoted to local variable blocks.
\begin{lemmax}[local-default-rely]
\(
 \Local{x}{c} = \Local{x}{c \together \rely{(\id(x))}} 
\)
\end{lemmax}

\begin{proof}
From the definition of localisation (\refdef{localise}) and
$\eguard{\id(x)} \together \rely{\id(x)}$
equals $\eguard{\id(x)}$ 
from
(\refprop{demand-rely}).
\end{proof}
\begin{lemmax}[variable-default-rely]
\(
  \variable{x}{c} = \variable{x}{c \together \rely{\id(x)}} 
\)
\end{lemmax}

\begin{proof}
From the definition of localisation (\refdef{localise}) and
\reflem{local-default-rely}.
\end{proof}

\subsubsection{Local variable block introduction.}
  
The proviso for the following lemma means that the effect of command $c$ on
variables other than $x$ is independent of $x$, and that program steps
of $c$ do not modify $x$, but that it $x$ is otherwise unconstrained.
\pagebreak[3]
\begin{lemmax}[introduce-local]
If $c = \Hide{x}{c} \together \guar{\id(x)}$, then
\(
  c = \Local{x}{\Hide{x}{c}}~.
\)
\end{lemmax}

\begin{proof}
\begin{displaymath}
\Local{x}{\Hide{x}{c}}
\Equals*[expand the definition of localisation (\refdef{localise})]
\Hide{x}{(\eguard{\id(x)} \together (\Hide{x}{c}))} \together \guar{\id(x)}
\Equals*[\localise\ operand of $\together$ (\refax{hide-distrib-sync})]
\Hide{x}{(\eguard{\id(x)})} \together (\Hide{x}{c}) \together \guar{\id(x)}
\Equals*[by assumption $c = \Hide{x}{c} \together \guar{\id(x)}$]
\Hide{x}{(\eguard{\id(x)})} \together c
\Equals*[by (\refprop{hide-demand-id-x}), $\Hide{x}{(\eguard{\id(x)})} = \Chaos$ which is the identity of $\together$]
c
\end{displaymath}
\vspace{-5ex}\\
\end{proof}

From \reflem{introduce-local} we have the following variation if the
condition on $c$ is a refinement instead of an equality.
\begin{lemmax}[introduce-local-refine]
If $c \refsto \Hide{x}{c} \together \guar{\id(x)}$,~~ 
\begin{math}
c \refsto \Local{x}{\Hide{x}{c}} .
\end{math}
\end{lemmax}

\begin{proof}
\begin{displaymath}
c \Refsto*[by assumption $c \refsto \Hide{x}{c} \together \guar{\id(x)}$]
\Hide{x}{c} \together \guar{\id(x)}
\Equals*[by \reflem{introduce-local}]
\Local{x}{\Hide{x}{(\Hide{x}{c} \together \guar{\id(x)})}}
\Equals*[by (\refax{hide-distrib-sync})]
\Local{x}{\Hide{x}{c} \together \Hide{x}{(\guar{\id(x)})}}
\Equals*[by (\refprop{hide-guar-id-x}), $\Hide{x}{\guar{\id(x)}} = \Chaos$ which is the identity of $\together$]
\Local{x}{\Hide{x}{c}}
\end{displaymath} 
\vspace{-5ex}\\
\end{proof}

The first proviso for introducing a local variable block ensures that
$c$ is not affected by any stuttering steps taken to allocate or
deallocate the variable. The second proviso is the same as for
localising the variable in \reflem{introduce-local}.
\begin{lawx}[introduce-variable]
Property $c = \variable{x}{\Hide{x}{c}}$ holds if both,
\begin{eqnarray*}
c & = & \Idle \Seq c \Seq \Idle \\
c & = & \Hide{x}{c} \together \guar{\id(x)} 
\end{eqnarray*}
\end{lawx}

\begin{proof}
\begin{displaymath}
c
\Equals*[using the first assumption]
\Idle \Seq c \Seq \Idle
\Equals*[using second assumption and \reflem{introduce-local}]
\Idle \Seq \Local{x}{\Hide{x}{c}} \Seq \Idle
\Equals*[by the local variable block definition (\refdef{var-intro})]
\variable{x}{\Hide{x}{c}}
\end{displaymath}
\vspace{-3ex}
\end{proof}

Similarly, if we have our conditions on $c$ are refinements instead of
equalities the following law applies from \Law{introduce-variable}.
\begin{lawx}[introduce-variable-refine]
If both, $c  \refsto  \Idle \Seq c \Seq \Idle$ and $c  \refsto  \Hide{x}{c} \together \guar{\id(x)}$, 
then
\begin{math}
c \refsto \variable{x}{\Hide{x}{c}} .
\end{math}
\end{lawx}

\subsubsection{Distributivity over local variable blocks.}

Commands that are \localised\ in variable $x$ distribute in and out of a
local scope $\Local{x}{c}$.

\begin{lemmax}[local-distribute]
If $\Hide{x}{d} = d$,
\begin{eqnarray*}
  \Local{x}{c} \together d = \Local{x}{c \together d}
\end{eqnarray*}
\end{lemmax}

\begin{proof}
The proof follows directly from the definition of localisation
(\refdef{localise}), and (\refax{hide-distrib-sync}) applied to
operation $\together$.
\end{proof}

For instance, using \reflem{localise-gdr}, we have that \localised\ 
rely and guarantee commands distribute in and out of local scopes,
e.g.
\begin{eqnarray}
  \Local{x}{c} \together \guar{g} & = & \Local{x}{c \together \guar{g}}, ~~~~~\mbox{if $g = \Hide{x}{g}$} \labelprop{local-guar-distribute}\\
  \Local{x}{c} \together \rely{r} & = & \Local{x}{c \together \rely{r}}, ~~~~~~~~\mbox{if $r = \Hide{x}{r}$}
\labelprop{local-rely-distribute}
\end{eqnarray}
Local variable blocks include idle steps before and after the
introduction of their local variable scope, and so distributivity of
\localised\ guarantees (\refprop{local-guar-distribute}) can be promoted
to local variable blocks provided the guarantee is reflexive, i.e. so
that $\Idle \together (\guar{g}) = \Idle$.
\begin{lawx}[variable-guarantee-distribute]
For 
\ihsb{R4 why reflexive? -- see 3rd step}{reflexive} relation $g$ such that $g = \Hide{x}{g}$,
\begin{eqnarray*}
  \variable{x}{c} \together \guar{g} = \variable{x}{c \together \guar{g}}
\end{eqnarray*}
\end{lawx}

\begin{proof}
\begin{displaymath}
\variable{x}{c} \together \guar{g}
\Equals*[expand local variable block definition (\refdef{var-intro})]
(\Idle \Seq \Local{x}{c} \Seq \Idle) \together \guar{g}
\Equals*[distribute guarantee over sequential composition by (\refprop{guar-distrib-seq})]
(\Idle \together \guar{g}) \Seq (\Local{x}{c} \together \guar{g}) \Seq (\Idle \together \guar{g})
\Equals*[by reflexivity of $g$ and (\refprop{idle-together-reflexive-guarantee})]
\Idle \Seq (\Local{x}{c} \together \guar{g}) \Seq \Idle
\Equals*[using assumption $g = \Hide{x}{g}$, \reflem{localise-gdr} and \reflem*{local-distribute}]
\Idle \Seq \Local{x}{c \together \guar{g}} \Seq \Idle
\Equals*[fold local variable block definition (\refdef{var-intro})]
\variable{x}{c \together \guar{g}}
\end{displaymath}
\end{proof}

From \Law*{variable-guarantee-distribute} and
\reflem{variable-default-guarantee} we can prove the following law
that was stated, but not proved in our semantics paper
\cite{DaSMfaWSLwC}. It states that any guarantee within a local
variable block on $x$ holds for the whole command provided that $x$ is
\localised\ from the guarantee. We can strengthen this guarantee with
the fact that any non-local occurrence of $x$ is not modified by the
local variable block:
\begin{lawx}[variable-guarantee]
For 
reflexive relation $g$,
\begin{eqnarray*}
\variable{x}{c \together \guar{g}} & = &
\variable{x}{c \together \guar{g}} \together \guar{((\Hide{x}{g}) \inter \id(x))}
\end{eqnarray*}
\end{lawx}

\begin{proof}
\begin{displaymath}
\variable{x}{c \together \guar{g}}
\Equals*[$g \subseteq (\Hide{x}{g})$ by (\refax{hide-weakens})]
\variable{x}{c \together \guar{(g \cap \Hide{x}{g})}}
\Equals*[splitting guarantee using (\refprop{merge-guarantees})]
\variable{x}{c \together \guar{g} \together \guar{(\Hide{x}{g})}}
\Equals*[by \Law*{variable-guarantee-distribute} and \reflem{variable-default-guarantee}]
\variable{x}{c \together \guar{g}} \together \guar{(\Hide{x}{g})} \together \guar{\id(x)}
\Equals*[merging guarantees by (\refprop{merge-guarantees})]
\variable{x}{c \together \guar{g}} \together \guar{((\Hide{x}{g}) \inter \id(x))}
\end{displaymath}
\vspace{-5ex}\\
\end{proof}

By (\refprop{weaken-rely}), weakening a rely condition produces a refinement, 
and so rely conditions can distribute, by
refinement, into local variable blocks in the following way.
\begin{lawx}[variable-rely-distribute-refine]
\begin{eqnarray*}
\variable{x}{c} \together \rely{r} & \refsto &  \variable{x}{c \together \rely{(\Hide{x}{r})}}
\end{eqnarray*}
\end{lawx}

\begin{proof}
\begin{displaymath}
\variable{x}{c} \together \rely{r}
\Equals*[expand local variable block definition (\refdef{var-intro})]
(\Idle \Seq \Local{x}{c} \Seq \Idle) \together \rely{r}
\Refsto*[by (\refprop{rely-distrib-seq}) twice]
(\Idle \together \rely{r}) \Seq
(\Local{x}{c} \together \rely{r}) \Seq
(\Idle \together \rely{r})
\Refsto*[using remove rely (\refprop{remove-rely}) twice, and weaken rely (\refprop{weaken-rely}) using $r \subseteq \Hide{x}{r}$]
\Idle \Seq
(\Local{x}{c} \together \rely{(\Hide{x}{r})}) \Seq
\Idle
\Equals*[\reflem{localise-gdr} and \reflem{local-distribute}]
\Idle \Seq
\Local{x}{c \together \rely{(\Hide{x}{r})}} \Seq
\Idle
\Equals*[fold local variable block definition (\refdef{var-intro})]
\variable{x}{c \together \rely{(\Hide{x}{r})}}
\end{displaymath}
\vspace{-5ex}\\
\end{proof}

Using these lemmas, we can verify the following law that was stated,
but not proved in \cite{DaSMfaWSLwC}:

\begin{lawx}[variable-distribute-rely]
\begin{eqnarray*}
\variable{x}{c} \together \rely{r} & \refsto & 
\variable{x}{c \together \rely{((\Hide{x}{r}) \inter \id(x))}}
\end{eqnarray*}
\end{lawx}

\begin{proof}
By \Law{variable-rely-distribute-refine},
\reflem{variable-default-rely}, and (\refprop{merge-relies}).
\end{proof}

\section{Related work and conclusions}

Earlier research on formalising rely/guarantee concurrency \cite{XuRoeverHe97,PrensaNieto03,CoJo07}
makes use of an operational semantics for reasoning about the correctness in
a Hoare logic style extended with rely and guarantee conditions
but these works do not consider local variables.
Concurrent Kleene Algebra (CKA) \cite{DBLP:journals/jlp/HoareMSW11} and 
Synchronous Kleene Algebra (SKA) \cite{Pris10} provide algebras for reasoning 
about concurrent programs in a style similar to that proposed here,
however, neither treat local variable blocks.

Dingel \cite{Dingel02} considers a refinement calculus for rely/guarantee concurrency
that uses a trace-based semantics to define his concurrent wide-spectrum language.
His language includes a local variable construct.
To express refinements involving local variables, 
he makes use of a refinement relation $C \succ_{V} C'$,
that states that $C$ is refined by $C'$ ignoring the values of (local) variables in $V$.
However, his rule for introducing a local variable $x$ requires 
that $x \notin V$, i.e.~$x$ is not already a local variable.
Such an approach is problematic for handling recursive procedures that include local variables
because the recursive calls use a local variable with the same name.
As Dingel does not handle procedures, this is not a problem in his context.
Our approach does not need a refinement relation dependent on a set of variables
and allows introducing a local variable with the same name as an existing variable.

Our experience with developing a concurrent refinement algebra [5,6] has shown that one can build a rich theory capable of supporting reasoning about concurrent programs in a rely/guarantee style starting from a small core theory of primitive commands and operators with simple algebraic properties. We have followed the same approach to adding local variables to our concurrent refinement algebra. Rather than local variable blocks being a primitive, we start with a much simpler localisation operator whose algebra is a variant of Tarski's cylindric algebra (for first order logic). It shares the same basic properties as Tarski's algebra but is extended to handle localisation of commands and operators on commands including sequential composition and weak conjunction.

In extending localisation to relations and commands, 
we found it necessary to introduce distinct localisations for the first, last and all states
in order to prove general properties about localisations of sequential compositions.
From the basic theory we are then able to prove properties about 
the interactions of localisations with constructs in our wide spectrum language,
such as relies and guarantees,
and hence, because local variable blocks are defined in terms of our localisation primitive,
we can also prove properties of interactions between local variable blocks and 
other constructs.

As well as the work presented here, we have examined the laws for localisation of
an assert command, $\Pre{p}$, and a specification command, $\Post{q}$,
however, these lemmas are not included due to page limitations.
Future work involves expanding the use of localisation to value parameters to procedures
and for handling data refinement, 
where the distinct first-state and last-state localisations are also a necessity.

\bibliographystyle{plain}
\bibliography{ms}

\end{document}